\documentclass[twocolumn,tighten]{aastex631}

\begin{document}

\shortauthors{Luhman}

\shorttitle{Membership and Age of IRAS 04125+2902}

\title{The Membership and Age of the Planet-hosting Young Star 
IRAS 04125+2902\footnote{Based on observations made with the Gaia 
mission, the Two Micron All Sky Survey, the Wide-field Infrared Survey 
Explorer, and the NASA Infrared Telescope Facility.}}

\author{K. L. Luhman}
\affiliation{Department of Astronomy and Astrophysics,
The Pennsylvania State University, University Park, PA 16802, USA;
kll207@psu.edu}
\affiliation{Center for Exoplanets and Habitable Worlds, The
Pennsylvania State University, University Park, PA 16802, USA}

\begin{abstract}

A transiting planet was recently discovered around a star in the
Taurus star-forming region, IRAS 04125+2902, making it one of the youngest
known planets. The discovery paper cited two age estimates for IRAS 04125+2902, 
one based on a comparison to two sets of model isochrones in the
Hertzsprung-Russell (H-R) diagram and a second age reported by an earlier study 
for a putative population in Taurus that includes IRAS 04125+2902 (D4-North).
However, the model isochrones in question differ significantly for most
temperatures and luminosities of young low-mass stars, and do not reproduce 
the observed sequences for the TW Hya and 32 Ori associations (10 and 21 Myr).
Meanwhile, as found in my previous work,
D4-North is a collection of field stars and fragments of several distinct
Taurus groups and older associations, so its quoted age is not meaningful.
The true parent population for IRAS 04125+2902 is a small group that is 
$\sim$35 pc behind the L1495 and B209 clouds (B209N).
I have analyzed the age of B209N through a comparison to TW Hya and 32 Ori.
The M star sequences in the latter two associations have the same shapes,
but the sequence for B209N is flatter, indicating that $>$M4 stars at 
ages of $<$10 Myr fade more quickly than stars at earlier types and older ages. 
For the one member of B209N that is earlier than M4 (IRAS 04125+2902),
I estimate an age of 3.0$\pm$0.4 Myr based on its offsets from TW Hya and
32 Ori, which by happenstance is similar to the value derived through
the comparison to model isochrones.

\end{abstract}

\section{Introduction}
\label{sec:intro}

Over the last decade, the K2 mission \citep{how14} and 
the Transiting Exoplanet Survey Satellite \citep{ric15} have proven to be 
successful at identifying transiting planets around young stars 
in the solar neighborhood
\citep[$<$50 Myr,][]{dav16,dav19,new19,bou20,pla20,man22}.  \citet{bar24} 
(hereafter B24) recently extended this work to particularly young ages
through the discovery of a transiting planet around IRAS 04125+2902,
which is low-mass star in the Taurus star-forming region 
\citep{luh23tau} (hereafter L23).

\citet{ken94} identified IRAS 04125+2902 as a candidate member of 
Taurus based on its photometry from the Infrared Astronomical Satellite, which
suggested the presence of a circumstellar disk.
\citet{luh09} confirmed its nature as a young star through spectroscopy,
and found evidence for a large inner hole in its disk (a transitional disk)
in the spectral energy distribution measured with the Spitzer Space Telescope
\citep{wer04}. \citet{reb10} and \cite{her14} subsequently presented similar 
spectral classifications of the star. The star's disk has been studied in 
more detail through Spitzer spectroscopy \citep{fur11} and submillimeter 
imaging \citep{esp15}. \citet{luh09} noted a candidate companion 
(2MASS J04154269+2909558) at a separation of
$4\arcsec$ from IRAS 04125+2902 in images from Spitzer and ground-based
telescopes, which was spectroscopically classified as a young M6.5 object
\citep{luh17}, placing it near the hydrogen burning limit \citep{bar15}. 
Astrometry from the Gaia mission \citep{per01,deb12,gaia16b} has supported
the binarity of the pair, and has indicated that the system resides in a small
group (B209N) that is $\sim$35~pc behind the L1495 and B209 clouds in Taurus 
\citep[][L23]{luh18}.

Detections of planets in star-forming regions offer the potential for 
observing the earliest stages of planet formation.
A key factor in the interpretation of such data is the age of the host star
for a given planet. B24 described two age estimates for IRAS 04125+2902,
an age of 2.49 Myr reported for a putative population in Taurus that 
contains IRAS 04125+2902
\citep[D4-North,][]{kro21} (hereafter K21) and their own estimate of 3.3 Myr
based on a comparison of the luminosity and temperature of the star to model
isochrones. In this paper, I discuss the evidence that IRAS 04125+2902 is
a member of the B209N group within Taurus and I attempt to constrain
the age of that group through a comparison to the TW Hya association (TWA)
and the 32 Ori association.

\section{Strategy for Estimating the Age of IRAS 04125+2902}

The age of a low-mass star in a star-forming region can be estimated
through a comparison of its temperature and luminosity (or related
parameters like color and absolute magnitude) to the isochrones predicted 
by theoretical evolutionary models.  
However, temperature and luminosity 
estimates for young stars can have large uncertainties because of spots, 
accretion, extinction, and unresolved companions \citep[e.g.,][]{per24}.
The random errors introduced by those phenomena can be reduced by deriving the 
median age for all low-mass stars in a population and assuming that the stars 
are coeval (i.e., comparing the sequence of low-mass stars in the 
Hertzsprung-Russell (H-R) diagram to model 
isochrones)\footnote{All members of the parent group of IRAS 
04125+2902 are considered in my age analysis, but the final age estimate is
based on that star alone for reasons described in 
Section~\ref{sec:compareobs}.}.
That approach is still subject to systematic errors that vary from one
set of models to another, as illustrated in Section~\ref{sec:age}.
Model isochrones often imply younger ages than those found with other 
methods \citep{nay09,bel13}.
To avoid those systematic errors, one can compare the H-R diagram sequence 
for a given population to sequences for other associations to determine
relative ages. If some of those associations have age estimates based on
lithium depletion or kinematic analysis, an absolute age 
can be derived that has less dependence on model isochrones \citep{her15}. 
That is the strategy that I pursue for IRAS 04125+2902.

\section{Membership of IRAS 04125+2902}

\begin{figure}
\epsscale{1.2}
\plotone{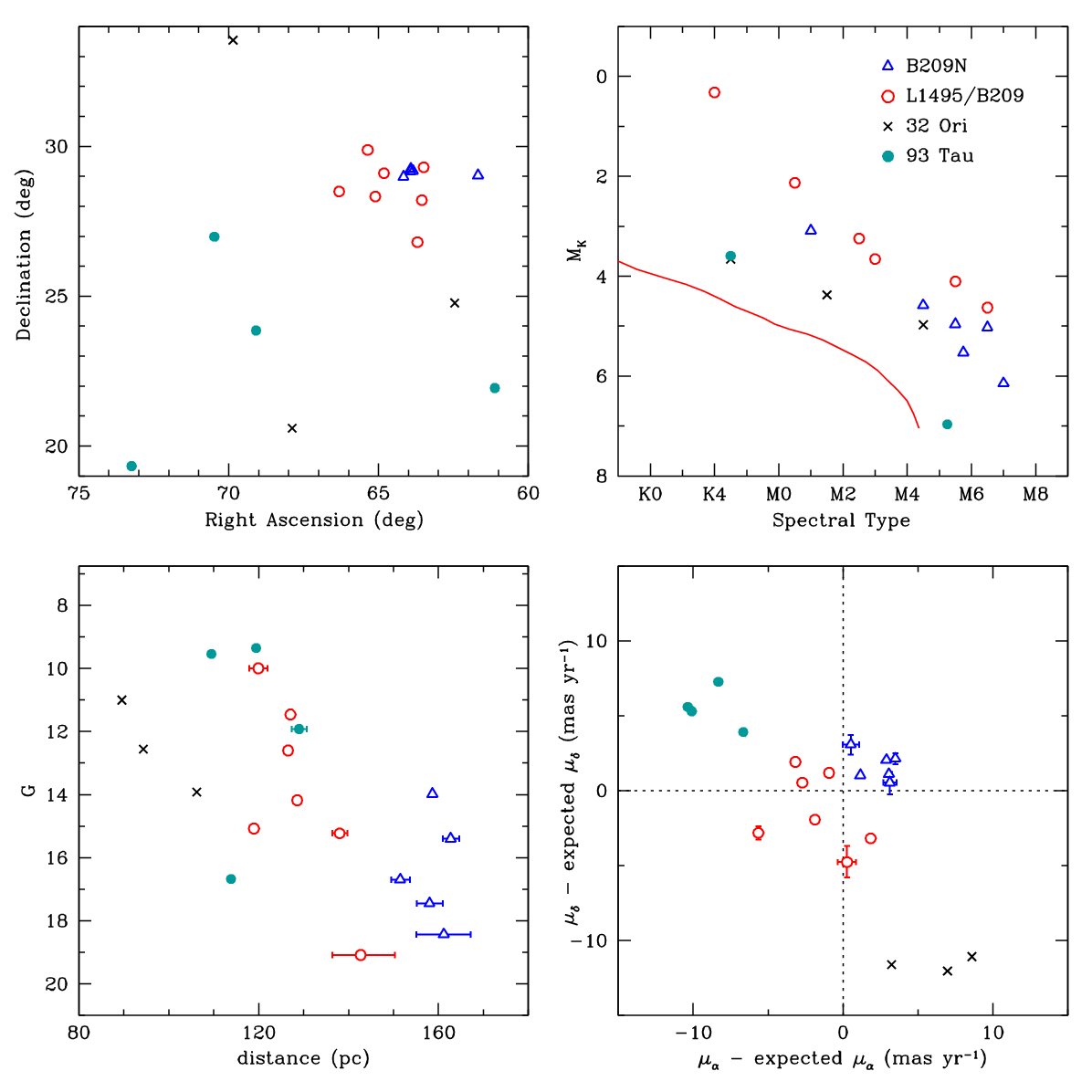}
\caption{
Equatorial coordinates, extinction-corrected $M_K$ versus spectral type,
$G$ versus parallax, and proper motion offsets for stars that were assigned
to the D4-North Taurus population by K21 and that 
\citet{luh18,luh22o,luh23tau} classified as members of two Taurus groups 
(B209N, L1495/B209) and two older associations in the vicinity of Taurus
(32 Ori, 93 Tau).  
A fit to the single-star sequence of the Pleiades is indicated
in the H-R diagram (red solid line).
Error bars are omitted for the distances and proper motion offsets
when they are smaller than the symbols.}
\label{fig:f1}
\end{figure}

K21 attempted to divide their compilation of adopted Taurus members into
separate populations by applying a Gaussian mixture model to the
Galactic Cartesian coordinates of those sources.  One of the resulting 
populations, D4-North, contains 83 stars, including IRAS 04125+2902.
K21 derived an age of 2.49 Myr for that sample through a comparison
to model isochrones, and that age was cited by B24 as applicable to
IRAS 04125+2902. However, in L23, I found that 
most of the populations from K21 are mixtures of multiple distinct groups,
and thus do not represent coherent populations. 
For instance, D4-North consists of fragments of several Taurus 
groups (as defined in L23), fragments of two older associations 
that are unrelated to Taurus (32 Ori and 93 Tau), and 19 field stars.
To illustrate those results, I have plotted equatorial coordinates,
an H-R diagram in the form of $M_K$ versus spectral type, $G$ versus distance, 
and proper motion offsets\footnote{Proper motion offset is defined as the 
difference between the observed proper motion of a star and the motion 
expected at the celestial coordinates and parallactic distance of the star 
for the median space velocity of Taurus.}
in Figure~\ref{fig:f1} for stars in D4-North that I classified as members 
of the L1495/B209 and B209N groups in Taurus, 32 Ori, and 93 Tau. 
The astrometry and $G$ measurements are from the third data release of Gaia 
\citep[DR3,][]{bro21,val23}, the distances are based on Gaia DR3 parallaxes
\citep{bai21}, and the $K_s$ photometry is from the Point Source Catalog
of the Two Micron All Sky Survey \citep[2MASS,][]{skr03,skr06}.
The four group fragments in Figure~\ref{fig:f1} have different kinematics and 
ages, indicating that they do not belong to a single population and that
the age reported for D4-North is not meaningful. Data like those in 
Figure~\ref{fig:f1} for the full samples of members of the Taurus groups
and the older associations were presented in L23.

\begin{figure}
\epsscale{1.2}
\plotone{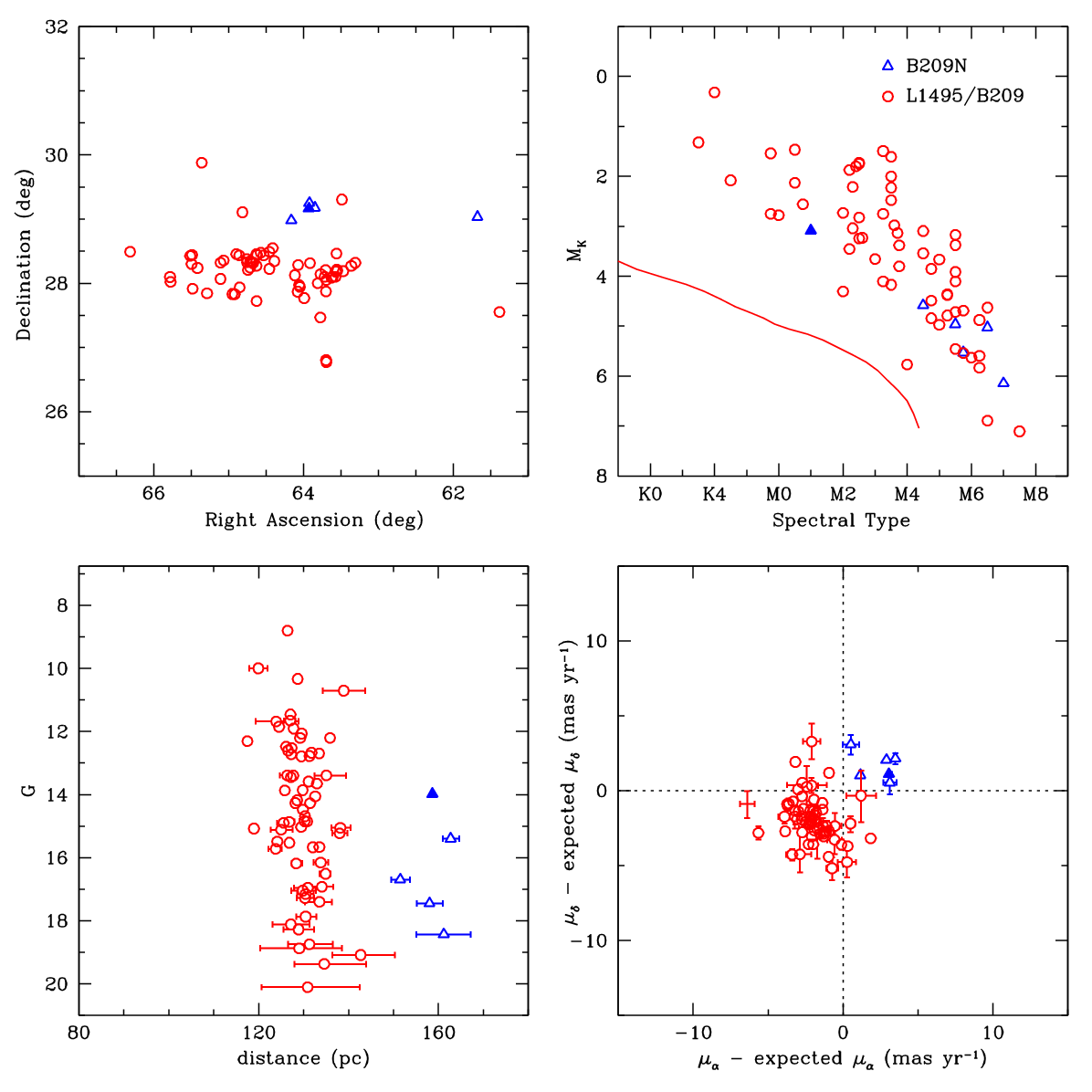}
\caption{
Equatorial coordinates, extinction-corrected $M_K$ versus spectral type,
$G$ versus parallax, and proper motion offsets for members of the B209N
and L1495/B209 groups in Taurus \citep{luh23tau}. The filled triangle
in each diagram is IRAS 04125+2902.}
\label{fig:f2}
\end{figure}

IRAS 04125+2902 is located $\sim1\arcdeg$ north of the L1495 and B209 
clouds in Taurus, so one might expect that it is a member of the group
of stars associated with those clouds. However, Gaia astrometry has 
demonstrated that it is a member of a small group that is spatially and 
kinematically distinct from the stars in L1495/B209 \citep{luh18}.
That group has six adopted members and was named B209N (L23).
IRAS 04125+2902 and its $4\arcsec$ companion are counted as
separate members. In Figure~\ref{fig:f2}, I present diagrams 
with equatorial coordinates, $M_K$ versus spectral type, $G$ versus distance,
and proper motion offsets for the members of L1495/B209 and B209N.
In addition to their different kinematics and spatial positions, the
two groups have different ages based on the H-R diagram.

As the brightest member of B209N, IRAS 04125+2902 has the smallest error
in its parallactic distance. Four of the five other members have separations of
$<1000\arcsec$ from IRAS 04125+2902, which correspond to $<0.8$ pc at
their distances. Since the fainter members have larger parallax
errors, I have adopted the distance of IRAS 04125+2902 for those four members
when calculating their $M_K$, which assumes that these stars are 
as tightly clustered in distance as they are on the sky. For the sixth
member (the westernmost one in Figure~\ref{fig:f2}), I have adopted its
parallactic distance, which has an error of 2 pc.

One member of B209N, Gaia DR3 164783811951433856,
has an uncertain parallax and a poor astrometric fit based on its 
renormalized unit weight error \citep[RUWE,][]{lin18}, but it is close to the
other members on the sky and has a similar proper motion. It is absent from
the diagram of $G$ versus distance in Figure~\ref{fig:f1} and \ref{fig:f2}. 
For the calculation of its proper motion offsets, I have adopted the distance 
of IRAS 04125+2902.

Some members of L1495/B209 lack measurements of proper motions
and parallaxes from Gaia DR3, so they are absent from the bottom
diagrams in Figure~\ref{fig:f2}. To plot those stars in the H-R diagram,
I have adopted the median distance of the members that do have parallax data.
The two class I protostars in L1495/B209 have been omitted from the
H-R diagram. In the diagrams of distance and proper motion offsets, I have
omitted a few members of L1495/B209 that have discrepant astrometry, as
noted in L23.

IRAS 04125+2902 has the earliest spectral type (M1) among the members of B209N.
The other members have types ranging from M4.5--M7, as shown in the 
H-R diagram in Figure~\ref{fig:f2}. The extinctions of these stars range
from $A_K=0.1$--0.3 \citep{esp19}. Some of that extinction may arise 
from the outskirts of the L1495 and B209 clouds, which are in the foreground 
of B209N.

\section{Age of the B209N Group}
\label{sec:age}

In L23, I estimated ages for some of the Taurus groups
based on the $M_K$ offsets of their sequences of low-mass stars (K4--M5) 
from the median sequence of Upper Centaurus-Lupus/Lower 
Centaurus-Crux (UCL/LCC), assuming that UCL/LCC has an age of 20~Myr 
\citep{luh22sc} and that the luminosities decrease at a rate given by
$\Delta$log~L/$\Delta$log~age$=-0.6$, which is the typical rate expected
from evolutionary models. Ages were not reported for the smallest
groups like B209N. Based on the two stars at K4--M5 in B209N (one of which
is IRAS 04125+2902), the implied age is 3.4 Myr. In this section, I attempt
a different approach that utilizes new age estimates for TWA and 32 Ori
that are based on expansion rate 
\citep[9.6$^{+0.9}_{-0.8}$ Myr,][]{luh23twa}\footnote{
\citet{mir25} reported that they measured radial velocities for 24 members of
TWA, which might enable refinement of the expansion age. However, they 
presented new measurements for only 14 stars, and only two of those velocities 
have significantly smaller errors than the values adopted in \citet{luh23twa}.
As a result, those new data have little effect on the estimated expansion age.}
and lithium depletion \citep[21.0$^{+1.0}_{-0.7}$ Myr,][]{luh24}, respectively.

\subsection{Construction of H-R Diagram}

My age analysis is performed with an H-R diagram that consists of 
$M_K$ versus spectral type. The $K_s$ band from 2MASS is selected because
it is long enough in wavelength that extinction is low while short enough
in wavelength that excess emission from disks is absent for most of the
stars in question.  Using this form of the H-R diagram, I will compare
B209N to TWA and 32 Ori because they are the youngest associations
that have well-defined sequences of low-mass stars and age estimates
from either expansion rate or lithium depletion.
I also include the Taurus group that contains HD 28354 because it
is between the ages of B209N and TWA and has a much tighter sequence than 
other Taurus groups (likely because of its older age).  
I omit stars that have full disks and that may have $K$-band excess emission.  
IRAS 04125+2902 lacks a $K$-band excess \citep{luh09}, and the
same is true for the remaining members of B209N.
I also omit stars that are blended with other sources in 2MASS.
For TWA and 32 Ori, I consider only the stars that are near the sequence 
of single stars for each association in Gaia color-magnitude diagrams 
\citep{luh22o,luh23twa}.

For B209N and the HD 28354 group, I use the spectral types adopted in L23,
all of which have been measured in my work or in a manner that is consistent 
with my classifications \citep{her14}.  
To check the previous optical classification 
of the $4\arcsec$ companion to IRAS 04125+2902, 2MASS J04154269+2909558, 
I have analyzed archival near-IR data (0.8--2.5~\micron, R=150)
that are available from SpeX \citep{ray03} at the NASA Infrared Telescope 
Facility, which were collected on 2017 February 15 through program 2017A104
(E. Magnier).  I reduced those data using the Spextool package \citep{cus04}, 
which included a correction for telluric absorption \citep{vac03}.
A comparison of the reduced spectrum to young standards \citep{luh17}
indicates a spectral type of M6.5, which is the same as the optical type.
The extinction produced by that comparison is $A_K=0.26$, which is similar
to the value for the primary \citep{esp19}. For other members of B209N
and HD 28354, I have adopted the extinction estimates from \citet{esp19}.
The extinction for IRAS 04125+2902 from that study agrees well with
the values derived by \citet{her14} and B24. In B209N, $M_K$ is calculated 
using the distances described in the previous section. For all other stars, 
the parallactic distances from \citet{bai21} are employed.
The sequences for TWA and 32 Ori are much tighter with Gaia colors than with 
spectral types, which reflects the smaller errors in the colors.
Therefore, I have converted $G_{BP}-G_{RP}$ to spectral type for each
star in TWA and 32 Ori using the relationship between spectral type and
$G_{\rm BP}-G_{\rm RP}$ for young stars from \citet{luh22sc}, assuming
that the stars have no extinction.

\subsection{Comparison of Model and Observed Isochrones}

\begin{figure}
\epsscale{1.2}
\plotone{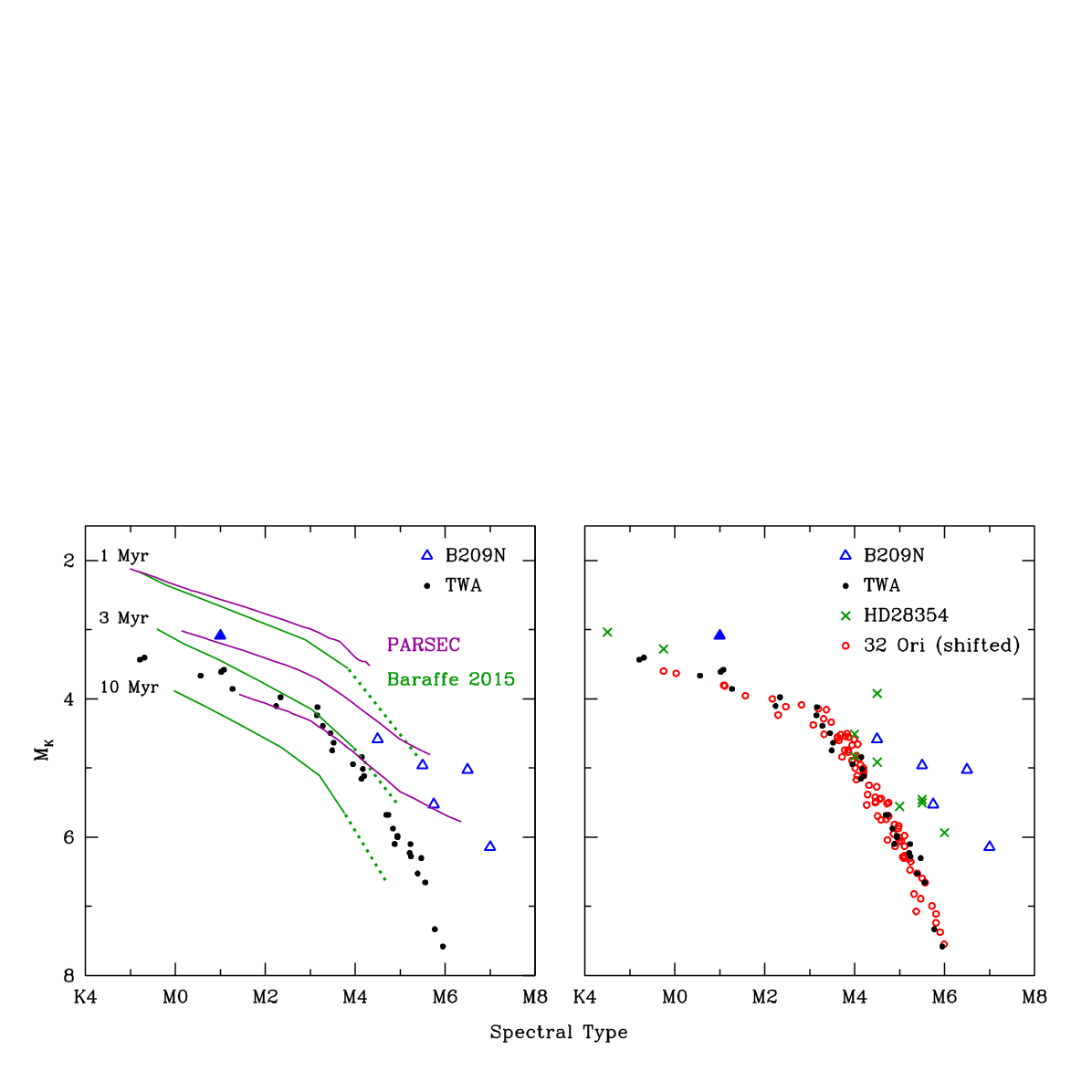}
\caption{
Left: $M_K$ versus spectral type for members of B209N and the single star
sequence of TWA compared to theoretical isochrones from \citet{bar15} 
(green lines) and the PARSEC models \citep{bre12,che14} (red lines).
The solid lines span masses of 0.2--0.7~$M_\odot$ and the dotted lines extend
to 0.1~$M_\odot$ for the Baraffe isochrones. Model temperatures have been
converted to spectral types using the temperature scale from \citet{her14}.
Right: B209N and TWA are compared to the HD~28354 group
in Taurus and the single star sequence of the 32 Ori association 
shifted to brighter $M_K$ by 0.42 mag.
Spectral types for TWA and 32 Ori have been derived from 
$G_{\rm BP}-G_{\rm RP}$ using the relationship between spectral type and 
$G_{\rm BP}-G_{\rm RP}$ for young stars \citep{luh22sc}.}
\label{fig:f3}
\end{figure}

To illustrate the variation among model isochrones for young low-mass
stars and compare them to observed sequences, I consider the isochrones
from \citet{bar15} and the PARSECv1.2 s isochrones \citep{bre12,che14},
which are the same sets of models that were utilized by B24 for estimating
the age of IRAS 04125+2902. The isochrones from \citet{bar15} are
similar to those from \citet{dot16} and \citet{cho16} for the masses and 
ages in question. I have converted the temperatures from the models to 
spectral types using the temperature scale for young stars from \citet{her14},
which is consistent with temperatures recently derived for K5--M2 stars in
Taurus \citep{per24}. In the left H-R diagram in Figure~\ref{fig:f3}, I have
plotted the Baraffe and PARSEC isochrones for masses of 0.2--0.7~$M_\odot$
and ages of 1, 3, and 10 Myr (solid lines). To show the shape of
the Baraffe isochrones down to M5--M6, I have extended them to
0.1~$M_\odot$ (dotted lines). For comparison to the models, 
I have plotted the members of B209N and the single star sequence of TWA.

As shown in the left diagram in Figure~\ref{fig:f3}, the Baraffe and PARSEC 
isochrones for young low-mass stars differ significantly, and neither set of 
models reproduces the sequence for TWA with the age derived from its expansion.
The PARSEC isochrone for 10 Myr is only slightly fainter than TWA
at M2-M3, but it differs increasingly at types later than M4.
Meanwhile, the Baraffe isochrone for 10 Myr better matches the shape of the 
TWA sequence, but it is too faint by nearly 1~mag. 
Figure~\ref{fig:f3} suggests that direct comparison of young low-mass stars
to model isochrones is unlikely to provide accurate ages. In the case of 
IRAS 04125+2902, B24 derived a temperature that corresponds
to M0 with the scale that I have adopted, resulting in a position in the
H-R diagram that is near the 3~Myr isochrones for both sets of models.
Based on the general disagreement between the two sets of isochrones,
I interpret the similarity of the ages that they produce for 
IRAS 04125+2902 as a coincidence rather an indication of their reliability.

\subsection{Comparison of Observed Sequences}
\label{sec:compareobs}

Members of the B209N and HD~28354 groups are plotted in the right
H-R diagram in Figure~\ref{fig:f3}.  The HD~28354 group has one bright
outlier (M4.5), which could be an unresolved binary.  The companion to 
IRAS 04125+2902 (M6.5) also may be overluminous relative to the remainder 
of the sequence for B209N. Otherwise, these groups have fairly well-defined
sequences that are much tighter than those of other Taurus groups (L23).

The right H-R diagram in Figure~\ref{fig:f3} includes the single
star sequences for TWA and 32 Ori. The shapes of those sequences are very
similar. To illustrate that, I have shifted 32 Ori to brighter magnitudes 
(by 0.42 mag) so that it is aligned with TWA. The fact that these sequences
are parallel indicates that stars across this range of temperatures fade
at the same rate between the ages of the two associations (10--21 Myr), 
which is expected from evolutionary models.
However, the B209N and HD~28354 groups are not parallel to TWA and 32 Ori,
and instead exhibit less curvature. For instance, the earliest
member of B209N, IRAS 04125+2902 (M1), is 0.62 mag brighter than TWA while
other members have progressively larger offsets at later spectral types,
ranging from 0.85--1.8 mag for M4.5--M5.75. The same trend is present in
the HD~28354 group, where the offsets are 0.24--1.7 mag for K8--M6. 
The change in the shape of these sequences between the Taurus groups
and TWA indicates that $>$M4 stars fade more rapidly than earlier stars
in that age range. Evolutionary models predict this behavior for the
least massive stars, although the change in the shape of the
isochrones (i.e., the fading rate for the least massive stars)
varies among different sets of models.

Given the expansion and lithium depletion ages for TWA and 32 Ori,
respectively, the offset of 0.42 mag between their sequences implies 
a fading rate of $\Delta$M$_K$/$\Delta$log~age$=1.24$, whereas 
a value near 1.5 is typically predicted by evolutionary models for the 
temperatures and ages in question.
Most evolutionary models predict that the fading rate for early-M stars
is roughly constant between ages of $\sim$1--100 Myr \citep{her15}.
Therefore, I have estimated the age of IRAS 04125+2902 from its 
offset from TWA (0.62 mag) by adopting the expansion age for TWA
and assuming that the fading rate between TWA and 32 Ori applies between
the ages of B209N and TWA, which results in an age of 3.0$\pm$0.4 Myr.
The error includes uncertainties in the expansion age of TWA, $K_s$
(0.03 mag), $A_K$ (0.03 mag), $M_K$ for the TWA sequence (0.02 mag), and the
spectral type (0.25 subclass).  This estimate is also subject to any
systematic errors in the adopted fading rate and the age for 32 Ori, which 
was derived with an empirical model for lithium depletion \citep{jef23}.
My analysis assumes that IRAS 04125+2902 does not have a close
binary companion that is unresolved by 2MASS, which is supported by
the high-resolution imaging and radial velocity measurements from B24.

Since my age estimate for IRAS 04125+2902 relies on photometry in $K_s$
from a single epoch, I have examined whether that measurement is likely to
be typical for the star. The best available constraints on the infrared
variability of IRAS 04125+2902 are from the Wide-field
Infrared Survey Explorer \citep[WISE,][]{wri10} and the reactivated mission
\citep[NEOWISE,][]{mai14}, which have provided all-sky images in the W1 and 
W2 bands (3.4 and 4.6~\micron) at multiple epochs between 2010 and 2024.
IRAS 04125+2902 was observed at 25 epochs, each of which consisted of
roughly a dozen exposures that spanned a few days. For each epoch, 
I have calculated the median values of the single-exposure measurements
in W1 and W2 from the AllWISE Multiepoch Photometry Table 
(\dataset[doi:10.26131/IRSA134]{http://dx.doi.org/doi:10.26131/IRSA134})
and the NEOWISE-R Single Exposure Source Table 
(\dataset[doi:10.26131/IRSA144]{http://dx.doi.org/doi:10.26131/IRSA144}).
The standard deviations of the medians from the 25 epochs are 0.017 and 0.015
mag in W1 and W2, respectively, indicating that the variability of
IRAS 04125+2902 is quite low. Its $4\arcsec$ companion is unresolved in the 
AllWISE and NEOWISE catalogs, but the remaining members of B209N also exhibit
low variability, which is likely one reason why B209N has a tighter sequence 
in the H-R diagram than other younger Taurus groups.

My age for IRAS 04125+2902 is similar to the value
of 3.3 Myr from B24, which was based on a comparison of the star to the PARSEC
isochrones. Although the methods for the two estimates differ, the resulting
ages are similar because of offsetting effects from (1) the
warmer temperature adopted by B24, (2) the PARSEC isochrones
(utilized by B24) appearing fainter than TWA near the temperature
of IRAS 04125+2902, and (3) the PARSEC models predicting a higher fading
rate than that derived from TWA and 32 Ori.

Ideally, the age of IRAS 04125+2902 would be further constrained through 
age estimates for the other members of B209N based on their offsets from TWA. 
However, all of the other members are in the range of spectral types ($>$M4)
where the shape of the sequence varies with age for stars younger than TWA,
reflecting a fading rate that varies with spectral type and age.
As a result, I cannot apply the fading rate between TWA and 32 Ori to
the late-type stars in B209N, and there is no empirical basis for
converting their offsets to ages.

\section{Conclusions}

A transiting planet was recently discovered around the low-mass star
IRAS 04125+2902 in the Taurus star-forming region (B24). 
I have presented an analysis of the membership and age of this star,
which is relevant to using the new companion to constrain the timing of
planet formation.

B24 described two age estimates for IRAS 04125+2902, one reported by K21 for a
putative population in Taurus that contains IRAS 04125+2902, which was named 
D4-North, and a second estimate based on a comparison of the luminosity and
temperature of the star to isochrones from \citet{bar15} and the PARSEC
models \citep{bre12,che14}.
However, as found in L23 and further demonstrated in this work, D4-North
consists of fragments of several distinct Taurus groups, fragments of
two older associations that are unrelated to Taurus (32 Ori and 93 Tau),
and 19 field stars. Thus, it is not a coherent population, and its quoted
age is not meaningful. Meanwhile, the Baraffe and PARSEC isochrones
for young low-mass stars differ significantly, and do not reproduce
the observed sequences for TWA and the 32 Ori association.
As a result, there no reason to expect that a direct comparison of the
luminosity and temperature of a young low-mass star like IRAS 04125+2902 to
model isochrones will provide a reliable age estimate.

Gaia astrometry has previously demonstrated that IRAS 04125+2902 is a member 
of a group of six stars that is 35~pc behind the L1495 and B209 clouds
in Taurus \citep{luh18}, which has the name of B209N (L23).
As illustrated in this work, B209N is spatially and kinematically distinct 
from the larger group of stars associated with the L1495 and B209 clouds,
and it has an older age based on the H-R diagram. 

I have attempted to estimate the age of B209N through a comparison to TWA and 
the 32 Ori association, which have recent age estimates of 10 and 21 Myr 
from expansion rate and lithium depletion, respectively. This comparison
is performed using the sequences of low-mass stars (M0--M6) in the H-R diagram.
The single star sequences for TWA and 32 Ori have the same shapes, which is
consistent with the predictions of evolutionary models for the ranges
of temperatures and ages in question.
However, the sequence for B209N is flatter than those of TWA and 32 Ori,
meaning that the offsets of B209N members above TWA and 32 Ori increase
with later spectral types. The same trend is found in a Taurus
group associated with HD 28354 that is between the ages of B209N and TWA.
The change in the shape of the sequences between Taurus and TWA indicates 
that $>$M4 stars fade more rapidly than earlier stars in that age range,
which is expected from evolutionary models.

I have derived a fading rate of $\Delta$M$_K$/$\Delta$log~age$=1.24$
between TWA and 32 Ori based on the offset between their sequences
and their expansion and lithium depletion ages, respectively. 
If I assume that rate is applicable between B209N and TWA for early-M stars,
then the offset of IRAS 04125+2902 (M1) above TWA implies an age of
3.0$\pm$0.4~Myr. All other members of B209N are cool enough ($>$M4) that
there is no empirical basis for converting their offsets to ages.
Although the methods differ, my age for IRAS 04125+2902 is similar to 
the value of 3.3 Myr derived by B24 because of multiple offsetting effects.

\begin{acknowledgments}

Gaia is a mission of the European Space Agency (\url{https://www.cosmos.esa.int/gaia}).
Its data have been processed by the Gaia Data Processing and Analysis 
Consortium (DPAC,
\url{https://www.cosmos.esa.int/web/gaia/dpac/consortium}). 
Funding for the DPAC has been provided by national institutions, in 
particular the institutions participating in the Gaia Multilateral Agreement. 
The IRTF is operated by the University of Hawaii under contract 80HQTR19D0030
with NASA. 2MASS is a joint project of the University of
Massachusetts and IPAC at Caltech, funded by NASA and the NSF.
WISE is a joint project of the University of California, Los Angeles,
and the JPL/Caltech, funded by NASA.
The Center for Exoplanets and Habitable Worlds is supported by the
Pennsylvania State University, the Eberly College of Science, and the
Pennsylvania Space Grant Consortium.

\end{acknowledgments}

\clearpage

\end{document}